\documentclass[12pt]{article}

\usepackage{color}
\usepackage{a4}
\usepackage{latexsym}
\usepackage{cite}
\usepackage{lineno}

\usepackage{array}
\newcolumntype{T}{>{\ttfamily} c}
\newcolumntype{M}{>{$\displaystyle} c <{$}}

\usepackage{epsfig}
\usepackage{graphicx}

\usepackage{pslatex}
\usepackage[latin1]{inputenc}
\usepackage[T1]{fontenc}

\usepackage{color,colordvi}
\def\colour4colour#1{\Blue{#1}}
\newcommand{\colourcolour}[1]{{\color{blue}{#1}}}

\newcommand{\gsim}{\raisebox{-0.7mm}{$\;\stackrel{>}{{\scriptstyle
 \sim}}\: $} }
\newcommand{\lsim}{\raisebox{-0.7mm}{$\:\stackrel{<}{{\scriptstyle
 \sim}}\: $} }

\newcommand{\beq}{\begin{equation}}
\newcommand{\eeq}{\end{equation}}
\newcommand{\bea}{\begin{eqnarray}}
\newcommand{\eea}{\end{eqnarray}}
\newcommand{\nn}{\nonumber}

\newcommand{\ra}{\rightarrow}
\newcommand{\equal}{\:\: = \:\:}

\newcommand{\als}{\alpha_{\rm s}}
\newcommand{\ars}{a_{\rm s}}
\newcommand{\ep}{\varepsilon}

\newcommand{\hspn}{{\hspace{-2mm}}}
\newcommand{\hspp}{{\hspace{3mm}}}

\def\frct#1#2{\mbox{\small{$\displaystyle\frac{#1}{#2}$}}}

\def\as(#1){{\alpha_{\rm s}^{\,#1}}}
\def\ar(#1){{a_{\rm s}^{\,#1}}}

\def\zr#1{{\zeta_{\:\!#1}^{}}}
\def\mus{{\mu^{\,2}}}

\def\B(#1,#2){{\beta_{#1}^{\,#2}}}

\def\nc{{n_c}}
\def\ncs{{n_{c}^{\,2}}}

\def\ncf{{n_{c}^{\,4}}}

\def\ca{{C^{}_A}}
\def\cas{{C^{\,2}_A}}
\def\cat{{C^{\,3}_A}}

\def\cf{{C^{}_F}}
\def\cfs{{C^{\, 2}_F}}
\def\cft{{C^{\, 3}_F}}

\def\nf{{n^{}_{\! f}}}

\def\nfs{{n^{\,2}_{\! f}}}
\def\nft{{n^{\,3}_{\! f}}}

\def\dFRAna{{ {d_{\,R}^{\,abcd}\,d_{\,A}^{\,abcd} \over n_a} }}
\def\dFRRna{{ {d_{\,R}^{\,abcd}\,d_{\,R}^{\,abcd} \over n_a} }}

\def\dfRAna{{ {d^{\,(4)}_{RA} \over n_a} }}
\def\dfRRna{{ {d^{\,(4)}_{RA} \over n_a} }}

\def\dFRAnA{{ {d_{\,R}^{\,abcd}\,d_{\,A}^{\,abcd} / n_a} }}
\def\dFRRnA{{ {d_{\,R}^{\,abcd}\,d_{\,R}^{\,abcd} / n_a} }}

\def\dfRAnA{{ {d^{\,(4)}_{RA}/n_a^{}} }}
\def\dfRRnA{{ {d^{\,(4)}_{RR}/n_a^{}} }}

\def\DNn#1{D_0^{\:#1}}
\def\DNm#1{D_{-1}^{\:#1}}
\def\DNp#1{D_1^{\:#1}}
\def\DNpp#1{D_2^{\:#1}}

\def\gqg0N{p_{\rm qg}^{}}

\def\xm1{{(1 \! - \! x)}}
\def\xp1{{(1 \! + \! x)}}

\def\Lnt(#1){\ln^{\,#1}(1\!-\!x)}
\def\pqq(#1){p_{\rm{qq}}(#1)}

\def\S(#1){{{S}_{#1}}}
\def\Ss(#1,#2){{{S}_{#1,#2}}}
\def\Sss(#1,#2,#3){{{S}_{#1,#2,#3}}}
\def\Ssss(#1,#2,#3,#4){{{S}_{#1,#2,#3,#4}}}
\def\Sssss(#1,#2,#3,#4,#5){{{S}_{#1,#2,#3,#4,#5}}}
\def\Ssssss(#1,#2,#3,#4,#5,#6){{{S}_{#1,#2,#3,#4,#5,#6}}}
\def\Sssssss(#1,#2,#3,#4,#5,#6,#7){{{S}_{#1,#2,#3,#4,#5,#6,#7}}}

\def\Sp(#1,#2){{{S}_{#1}^{\,#2}}}

\def\H(#1){{\rm{H}}_{#1}}
\def\Hh(#1,#2){{\rm{H}}_{#1,#2}}
\def\Hhh(#1,#2,#3){{\rm{H}}_{#1,#2,#3}}
\def\Hhhh(#1,#2,#3,#4){{\rm{H}}_{#1,#2,#3,#4}}
\def\Hhhhh(#1,#2,#3,#4,#5){{\rm{H}}_{#1,#2,#3,#4,#5}}
\def\Hhhhhh(#1,#2,#3,#4,#5,#6){{\rm{H}}_{#1,#2,#3,#4,#5,#6}}

\begin{document}
\setlength{\parskip}{0.2cm}
\setlength{\baselineskip}{0.54cm}

% --------------------------------------------------------------------

\begin{titlepage}
\noindent
DESY 23--096 \hfill July 2023\\
LTH 1345 \\
\vspace{0.6cm}
\begin{center}
{\LARGE \bf Four-loop splitting functions in QCD \\[1ex]
  -- The gluon-to-quark case --}\\ 
\vspace{2.0cm}
\large
G.~Falcioni$^{\, a}$, F.~Herzog$^{\, a}$, S. Moch$^{\, b}$ and A. Vogt$^{\, c}$\\

\vspace{1.2cm}
\normalsize
{\it $^a$Higgs Centre for Theoretical Physics, School of Physics and Astronomy\\
  The University of Edinburgh, Edinburgh EH9 3FD, Scotland, UK}\\
\vspace{5mm}
\normalsize
{\it $^b$II.~Institute for Theoretical Physics, Hamburg University\\
\vspace{0.5mm}
Luruper Chaussee 149, D-22761 Hamburg, Germany}\\
\vspace{5mm}
{\it $^c$Department of Mathematical Sciences, University of Liverpool\\
\vspace{0.5mm}
Liverpool L69 3BX, United Kingdom}\\
\vspace{3cm}
{\large \bf Abstract}
\vspace{-0.2cm}
\end{center}
We have computed the even-$N$ moments $N \leq 20$ of the gluon-to-quark
splitting function $P_{\rm qg}$ at the fourth order of perturbative QCD 
via the renormalization of off-shell operator matrix elements.
Our results, derived analytically for a general gauge group, agree with 
all results obtained for this function so far, in particular with the 
lowest five moments obtained via physical cross sections.
Using our new moments and all available endpoint constraints, we 
construct approximations for the four-loop $P_{\rm qg}(x)$ that should 
be sufficient for a wide range of collider-physics applications.
The~N$^3$LO corrections resulting from these and the corresponding
quark-quark splitting functions lead to a marked improvement of the
perturbative accuracy for the scale derivative of the singlet quark 
distribution, with effects of 1\% or less at $x \gsim 10^{\,-4}$ at a 
standard reference scale with $\als = 0.2$.
\vspace*{0.5cm}
\end{titlepage}

\newpage
% --------------------------------------------------------------------

High-energy particle physics is entering an era of precision measurements. 
Systematic errors at the ATLAS and CMS experiments at the Large Hadron 
Collider (LHC) will shrink towards the percent level 
\cite{ATLAS:2022hro,CMS:2021xjt}. 
Statistical uncertainties of many observables will be drastically reduced 
by increasing the amount of collected data by a factor of about 20 during 
the high-luminosity phase of the LHC \cite{Dainese:2019rgk}. 
Furthermore, new measurements of deep-inelastic scattering (DIS) with 
percent-level precision are expected at the forthcoming Electron-Ion 
Collider (EIC) \cite{AbdulKhalek:2021gbh}.

In order to match this accuracy in theoretical predictions of hard LHC 
processes and DIS, it is mandatory to compute radiative corrections up to 
the next-to-next-to-next-to-leading order (N$^3$LO) of perturbative QCD 
\cite{Caola:2022ayt}. 
The partonic cross sections for some key processes at the LHC
\cite{Anastasiou:2015vya,Mistlberger:2018etf,Duhr:2020seh,Chen:2021isd,%
Baglio:2022wzu} and for the structure functions $F_{1,2,3}$ in inclusive 
DIS \cite{Vermaseren:2005qc,Moch:2008fj,Davies:2016ruz} are already 
available at this accuracy.
% in the framework of collinear factorization. 
The N$^3$LO (four-loop) contributions to the splitting functions 
governing the scale evolution of the parton distribution functions (PDFs) 
are needed to complete these results.

Approximate results for the corresponding four-loop quark-quark splitting 
functions, which should be sufficient for most collider-physics 
applications, have been obtained for the non-singlet cases in 
refs.~\cite{Moch:2017uml,MVV-tba} and, recently, for the pure-singlet (ps) 
case in ref.~\cite{Falcioni:2023luc}. 
In this letter, we continue to address the scale dependence of the 
(unpolarized) flavour-singlet PDFs,
\beq
\label{eq:sgEvol}
  \frac{d}{d \ln\mus} \;
  \Big( \begin{array}{c} \! q_{\rm s}^{} \!\! \\ \!g\!  \end{array} \Big)
  \: = \: \left(
  \begin{array}{cc} \! P_{\rm qq} & P_{\rm qg} \!\!\! \\
                    \! P_{\rm gq} & P_{\rm gg} \!\!\! \end{array} \right)
  \otimes
  \Big( \begin{array}{c} \!q_{\rm s}^{}\!\! \\ \!g\!  \end{array} \Big)
  \: .
\eeq
Here 
$q_{\rm s}^{} \,=\, \sum_{\,i=1}^{\,\nf} \, ( q_i^{} + \bar{q}_i^{} )$ and
$g$ are the singlet quark and gluon distributions, with $\nf$ the number
of light flavours, and $\otimes$ denoting the Mellin convolution in the 
momentum variable $x$.

The splitting functions $P_{\,\rm{ij}}$, and the corresponding anomalous
dimensions $\gamma_{\,\rm ij}$ related to their even-$N$ Mellin moments by a 
conventional sign, can be expanded in powers of $\ars=\als(\mus)/(4\pi)$,
\beq
\label{eq:Pgamma}
  P_{\rm ij}^{}(x,\als) \:=\: \sum_{n=0} \ar(n+1)\,P_{\rm ij}^{\,(n)}(x)
\:\: , \quad
  \gamma_{\,\rm ij}^{\,(n)}(N) \; = \; - \int_0^1 \!dx\:\, x^{\,N-1}\,
  P_{\,\rm ij}^{\,(n)}(x)
\; .
\eeq
The next-to-next-to leading order (NNLO, N$^2$LO) $n=2$ terms in 
eq.~(\ref{eq:Pgamma}) were computed almost 20 years ago
\cite{Moch:2004pa,Vogt:2004mw}.
Over the years these were checked bit by bit in
various calculations, recently a complete re-calculation was performed 
using the operator-product expansion (OPE) \cite{Gehrmann:2023ksf}. 

Here we derive, in the same approach, the first 10 even-$N$ 
values of the anomalous dimensions $\gamma_{\,\rm qg}^{\,(3)}$, thus 
completing the upper row of the N$^3$LO matrix in eq.~(\ref{eq:sgEvol}) 
to $N = 20$. We then combine these results with large-$x$ and small-$x$ 
constraints to provide approximate expressions for $P_{\rm qg}^{\,(3)}(x)$.

The quantities $\gamma_{\,\rm ij}$ are the anomalous dimensions of the 
gauge invariant twist-two operators 
\beq
  O_{\rm q}^{\{\mu^{\,}_1,\cdots,\mu^{\,}_N\}} \:=\:
  \frct{1}{2}\,\overline{\psi}\,
  \gamma^{\,\{\mu^{}_1}D^{\,\mu^{}_2}\ldots D^{\,\mu^{}_N\}}\,\psi
 \,+\, \ldots
\:\: , \;\;
  O_{\rm g}^{\{\mu^{\,}_1,...,\mu^{\,}_N\}} \:=\:
  \frct{1}{2}\,F^{\nu \{ \mu^{}_1}
  D^{\,\mu^{}_2}\cdots D^{\,\mu_{N-1}}\, F^{\mu^{}_N \}}_{\hspace*{4mm}
  \nu} \,+\, \ldots
\eeq
where $\psi$ is the quark field, $F^{\mu\nu}$ is the gluon field 
strength tensor, and $D^{\,\mu} = \partial^{\,\mu} - ig\,A^\mu$ is the 
covariant derivative with the coupling $g$, where $g^2/(4\pi) = \als$. 
The curly brackets indicate symmetrization in the indices 
$\mu_1^{} \dots \mu_N^{}$. For brevity, we have not written out the 
additional terms that render these operators traceless.

$O_{\rm q,g}$ renormalize multiplicatively with a matrix of 
renormalization constants $Z_{\,\rm ij}$.
As already pointed out in ref.~\cite{Gross:1974cs}, $O_{\rm q}$ and 
$O_{\rm g}$ mix with a set of gauge-variant operators, known as 
aliens. The basis of aliens was determined in ref.~\cite{Dixon:1974ss} 
at two loops.
The general theory on the renormalization of gauge-invariant operators 
was then developed in refs.~\cite{Kluberg-Stern:1974nmx,%
Kluberg-Stern:1975ebk,Joglekar:1975nu,Joglekar:1976eb,Joglekar:1976pe}.
Recently the basis of alien operators up to four loops was constructed 
explicitly for any fixed moment $N$ \cite{Falcioni:2022fdm}. 
This basis is consistent with the counterterms computed at two and 
three loops for all values of $N$ in refs.~\cite{Dixon:1974ss,%
Hamberg:1991qt,Blumlein:2022ndg} and \cite{Gehrmann:2023ksf}, 
respectively. 
For the case at hand the renormalization proceeds along the lines of 
these references, but since it is technically more involved we defer 
the details to a later publication.

The anomalous dimensions $\gamma_{\,\rm ij}$ in eq.~(\ref{eq:Pgamma}) 
are obtained from the renormalization constants $Z_{\,\rm ij}$.
They are determined by requiring the finiteness of the renormalized 
operator matrix elements (OMEs) 
$A_{\rm ij} = \langle j(p) | O_{\rm i} | j(p)\rangle$. 
These are two-point Green functions of the operator $O_{\rm i}$ with 
off-shell external states $j$ of momentum $p$, where $j$ can be a quark 
($q$), gluon ($g$) or ghost~($c$). 

The Feynman diagrams for the OMEs have been generated using 
{\sc Qgraf}~\cite{Nogueira:1991ex}
and then processed, see ref.~\cite{Herzog:2016qas}, by a {\sc Form}
\cite{Vermaseren:2000nd,Kuipers:2012rf,Ruijl:2017dtg} program
which collects self-energy insertions, determines the colour factors 
\cite{vanRitbergen:1998pn} and classifies the topologies according to 
the conventions of the {\sc Forcer} program~\cite{Ruijl:2017cxj}.
An optimized in-house version of this program has been employed
to perform the integral reduction for fixed even values of $N$ in 
$4-2\ep$ dimensions.
Diagrams with the same colour factor and topology have been merged 
into meta-diagrams for computational efficiency.
 
In this manner, we were able to compute the physical OMEs $A_{\rm qg}$
for the gluon-to-quark splitting function to four loops for 
$N \leq 20$. The other physical OMEs $A_{\rm ps}$, $A_{\rm gq}$ and 
$A_{\rm gg}$ are needed at three loops for the determination of
$\gamma_{\,\rm qg}^{\,(3)}$, and the OMEs 
with the alien operators inserted into a gluon two-point function, 
$A_{\rm Ag}$, are required only to two loops.
These computations yield the following results for the N$^3$LO 
contributions to the anomalous dimensions $\gamma_{\,\rm qg}$ 
in eq.~(\ref{eq:Pgamma}) for QCD, i.e., the gauge group SU$(\nc=3)$:
\bea
\label{eq:gqg3-num}
  \gamma_{\,\rm qg}^{\,(3)}(N\!=\!2) \; & =\! & 
       - 654.4627782 \,\nf
       + 245.6106197 \,\nfs
       - 0.924990969 \,\nft
\, , \nn \\
  \gamma_{\,\rm qg}^{\,(3)}(N\!=\!4) \; & =\! & \phantom{-}
         290.3110686 \,\nf
       - 76.51672403 \,\nfs
       - 4.911625629 \,\nft
\, , \nn \\
  \gamma_{\,\rm qg}^{\,(3)}(N\!=\!6) \; & =\! & \phantom{-} 
         335.8008046 \,\nf
       - 124.5710225 \,\nfs
       - 4.193871425 \,\nft
\, , \nn \\
  \gamma_{\,\rm qg}^{\,(3)}(N\!=\!8) \; & =\! & \phantom{-}
         294.5876830 \,\nf
       - 135.3767647 \,\nfs
       - 3.609775642 \,\nft
\, , \nn \\
  \gamma_{\,\rm qg}^{\,(3)}(N\!=\!10) & =\! & \phantom{-}
         241.6153399 \,\nf
       - 135.1874247 \,\nfs
       - 3.189394834 \,\nft
\, , \nn \\
  \gamma_{\,\rm qg}^{\,(3)}(N\!=\!12) & =\! & \phantom{-}
         191.9712464 \,\nf
       - 131.1631663 \,\nfs
       - 2.877104430 \,\nft
\, , \nn \\
  \gamma_{\,\rm qg}^{\,(3)}(N\!=\!14) & =\! & \phantom{-}
         148.5682948 \,\nf
       - 125.8231081 \,\nfs
       - 2.635918561 \,\nft 
\, , \nn \\
  \gamma_{\,\rm qg}^{\,(3)}(N\!=\!16) & =\! & \phantom{-}
         111.3404252 \,\nf
       - 120.1681987 \,\nfs
       - 2.443379039 \,\nft
\, , \nn \\
  \gamma_{\,\rm qg}^{\,(3)}(N\!=\!18) & =\! & \phantom{-}
         79.51561588 \,\nf
       - 114.6171354 \,\nfs
       - 2.285486861 \,\nft 
\, , \nn \\
  \gamma_{\,\rm qg}^{\,(3)}(N\!=\!20) & =\! & \phantom{-}
         52.24329555 \,\nf
       - 109.3424891 \,\nfs
       - 2.153153725 \,\nft
\;.
\eea
The corresponding exact results, in terms of fractions and the values 
$\zr3$, $\zr4$ and $\zr5$ of the Riemann $\zeta$-function, are given for 
a general compact simple gauge group in app.~A,
eqs.~(\ref{eq:GqgN2}) -- (\ref{eq:GqgN20}).

Our results for $N \leq 10$ agree with refs.~\cite{Moch:2021qrk,MRUVV-tba},
where those moments were computed, as the three-loop all-$N$ expressions 
in refs.~\cite{Moch:2004pa,Vogt:2004mw}, via structure functions in 
inclusive DIS, a route that is conceptionally simpler but far more
demanding in terms of the integral reductions. The coefficients of $\nft$
in eq.~(\ref{eq:gqg3-num}) agree with the all-$N$ results in eq.~(3.12)
of ref.~\cite{Davies:2016jie}.
 
All-$N$ expressions for the anomalous dimensions include Riemann-$\zeta$ 
values, harmonic sums \cite{Vermaseren:1998uu} and simple denominators
 $D_a^{\,k} \equiv (N+a)^{-k}$. 
The latter frequently arise in the combinations
\bea
\label{eq:eta+nuDef}
  \eta &\!\equiv\!& \frct{1}{N} - \frct{1}{N+1} 
       \;=\; \DNn{} - \DNp{}
\:\: , \qquad
  \nu \;\equiv\; \frct{1}{N-1} - \frct{1}{N+2}
      \;=\; \DNm{} - \DNpp{}
\:\: , \\
\label{eq:gqpDef}
  \gqg0N \!\! &\!\equiv\!& 
  -\frct{2}{N} + \frct{4}{N+1} - \frct{4}{N+2}
      \;=\; -2\, \DNn{} + 4\DNp{} - 4\DNpp{}
\; .
\eea
Besides the leading large-$\nf$ contribution, all-$N$ expressions
for $\gamma_{\,\rm qg}^{\,(3)}$ have been obtained until now only 
for the $\zr4$ part, in eq.~(10) of ref.~\cite{Davies:2017hyl}, and the 
$\zr5$ coefficients of the quartic group invariants \cite{Moch:2018wjh}.
Using our results to $N=20$, we have now been able to determine the
$\zr5$ coefficients for all colour factors, and to extend the all-$N$
results for the quartic group invariants to the $\zr3$ terms.

The latter results, as in ref.~\cite{Moch:2018wjh} using the short-hand 
$ d^{\,(4)}_{xy} \;\equiv\; d_x^{\,abcd} d_y^{\,abcd} $,
are given by
\bea
  \label{gqg3z3d4RA}
\lefteqn{  \left. \gamma_{\,\rm qg}^{\,(3)}(N)
  \right|_{\,\zr3\, \colourcolour{\nf\*\dfRAnA}} \,=\;
%%START
%%L %%texgqg3z3d4RAna =
       128\,\*  \Big\{
       \Big( - {7/12} 
       - {33/2}\,\* \eta
       + 6\* \Big(16\*\eta - 16/3\*\nu + \eta^2\Big)\* \S(1)
%%STOP
}
%%START
   \nn \\[0mm] & & \mbox{} \vphantom{\Big(}
       + 2\* \Big(12 - 5\* \eta + 8\* \nu\Big)\* \S(-2)
       - \Big(43\* \eta - 16\* \nu \Big)\* \S(1)^2
       - 16\*\Ss(1,-2)
       + \S(3)
       + 8\*\S(-3)
     \Big) \* \gqg0N
   \nn \\[0mm] & & \mbox{} \vphantom{\Big(}
     + 4\* \Big( - 20 
       - 12\* \S(1) 
       + 36\* \S(-2) 
       + 5\* \S(1)^2 
       - 72\* \Ss(-2,1)
       - 18\* \S(3) 
       + 36\* \S(-3)
     \Big) \* D_{0}\*D_{1}\*D_{2}
    \Big\}
%%;
%%STOP
\; ,
\eea
\bea
  \label{gqg3z3d4RR}
\left. \gamma_{\,\rm qg}^{\,(3)}(N)
% \right|_{\,\zr3\, \colourcolour{\nfs\*\dfRRnA}} \;=\,
  \right|_{\,\zr3\, \colourcolour{\nfs\*\dfRRnA}} \,=\;
%%START
%%L %%texgqg3z3d4RRna =
  256\* \Big\{
     \Big( {7/12} 
       + 3\* \eta
       + 8\* \S(1)\* \eta^2 
       - 8\* \S(-2)\,\* \eta
     \Big) \* \gqg0N
     + 8\*\Big( 1
       - 4\* \S(-2)
     \Big) \* D_{0}\*D_{1}\*D_{2}
    \Big\}
%%;
%%STOP
\, ,
\eea
where the argument $N$ of the harmonic sums $S_{\dots}$ has been
suppressed. The complete $\zr5$ part reads
\bea
\label{gqg3z5}
\lefteqn{  \left. \gamma_{\,\rm qg}^{\,(3)}(N)
  \right|_{\,\zr5} \,=\;
%%START
%%L %%texgqg3z5 =
         \colourcolour{\nf\*\cft}\* \Big(
            {40/3}\,\*(N-1)
          - 640\*\gqg0N\*\S(1)
          - {41120/3}\,\*D_{1}
          - 7360\*D_{1}^2
          - 2560\*D_{2}
%%STOP
}
%%START
   \nn \\[0mm] & & \mbox{} \vphantom{\Big(}
          + {37120/3}\,\*D_{0}
          - 3680\*D_{0}^2
           \Big)
       + \colourcolour{\nf\*\cfs\*\ca}\*  \Big(
          - {80/3}\,\*(N-1)
          + 1200\*\gqg0N\*\S(1)
          - 640\*D_{-1}
   \nn \\[0mm] & & \mbox{} \vphantom{\Big(}
          + {65440/3}\,\*D_{1}
          + 16480\*D_{1}^2
          + 9760\*D_{2}
          + 1920\*D_{2}^2
          - {69800/3}\,\*D_{0}
          + 8240\*D_{0}^2
           \Big)
   \nn \\[0mm] & & \mbox{} \vphantom{\Big(}
       + \colourcolour{\nf\*\cf\*\cas}\*  \Big(
            {40/3}\,\*(N-1)
          - 1000\*\gqg0N\*\S(1)
          + 320\*D_{-1}
          - {40160/3}\,\*D_{1}
          - 10560\*D_{1}^2
          - 6480\*D_{2}
   \nn \\[0mm] & & \mbox{} \vphantom{\Big(}
          - 960\*D_{2}^2
          + {43840/3}\,\*D_{0}
          - 5280\*D_{0}^2
           \Big)
       + \colourcolour{\nf\*\cat}\*  \Big(
            {1000/3}\,\*\gqg0N\*\S(1)
          + {2240/9}\,\*D_{-1}
          + {31360/9}\,\*D_{1}
   \nn \\[0mm] & & \mbox{} \vphantom{\Big(}
          + 320\*D_{1}^2
          - {2960/3}\,\*D_{2}
          - {2240/3}\,\*D_{2}^2
          - {17840/9}\,\*D_{0}
          + 160\*D_{0}^2
           \Big)
       + \colourcolour{\nf\*\dfRAnA} \*  \Big(
          - 2560\*\gqg0N\*\S(1)
   \nn \\[0mm] & & \mbox{} \vphantom{\Big(}
          - {5120/3}\*D_{-1}
          - {129280/3}\,\*D_{1}
          - 26880\*D_{1}^2
          - 6400\*D_{2}
          + 5120\*D_{2}^2
          + {125120/3}\,\*D_{0}
   \nn \\[0mm] & & \mbox{} \vphantom{\Big(}
          - 13440\*D_{0}^2
           \Big)
       + \colourcolour{\nfs\*\cas}\* \Big(
            {5120/9}\,\*D_{1}
          + {1280/3}\,\*D_{1}^2
          + {640/9}\,\*D_{2}
          - {5440/9}\,\*D_{0}
          + {640/3}\,\*D_{0}^2
           \Big)
   \nn \\[0mm] & & \mbox{} \vphantom{\Big(}
       + \colourcolour{\nfs\*\dfRRnA} \*  \Big(
            {81920/3}\,\*D_{1}
          + 20480\*D_{1}^2
          + {10240/3}\,\*D_{2}
          - {87040/3}\,\*D_{0}
          + 10240\*D_{0}^2
           \Big)
%%;
%%STOP
\: .
\eea
Analytic expressions in $x$-space are, for now, available only for the 
leading large-$\nf$ part of $P_{\rm qg}^{\,(3)}$, see eq.~(4.22) of 
ref.~\cite{Davies:2016jie}. 
The above partial $N$-space results proportional to Riemann-$\zeta$ 
values do not translate to $x$-space Riemann-$\zeta$ expressions, 
since additional terms with $\zeta_n$ are generated by the inverse 
Mellin transformation.
Similarly, it is not possible to read off coefficients of $\zeta_n$ 
in the large-$N$ limit from eqs.~(\ref{gqg3z3d4RA}) -- (\ref{gqg3z5}).
In particular, as $\gamma_{\,\rm qg}$ vanishes for $N \ra \infty$, 
terms of the form $\zr5 \,(N-1)$ need to be compensated by other 
contributions that develop $\zr5$ terms in this limit, 
as in the functions $g_i(N)$ in eqs.~(3.18) -- (3.20) 
of ref.~\cite{Vermaseren:2005qc}.  

For the time being, only approximations can be provided for the N$^3$LO 
splitting function $P_{\rm ij}^{\,(3)}(x)$ in eqs.~(\ref{eq:sgEvol}) 
and (\ref{eq:Pgamma}), based on the moments (\ref{eq:gqg3-num}) 
and all known results for the large-$x$ and small-$x$ limits.
The large-$x$ expansion of $P_{\rm qg}^{\,(n)}(x)$ is given by
\beq
\label{eq:large-x}
  P_{{\rm qg},\, x \ra 1}^{\,(n)}(x) \; = \;
  \sum_{\ell\,=\,0}^{2n} \, \sum_{p\,=\,0}^{\infty} 
  C_{n,\ell,p}^{\:\rm qg} \, \xm1^p \, \ln^{\,2n-\ell\!} \xm1 
\; .
\eeq
The coefficients $C_{3,\ell,p}^{\:\rm qg}$ have been predicted
for $\ell = 0,\,1,\,2$ in ref.~\cite{Soar:2009yh}; the results for 
$p=0$ have been confirmed and extended to all higher orders $n$ 
in refs.~\cite{Vogt:2010cv,Almasy:2010wn}.
The small-$x$ expansion reads
\beq
\label{eq:small-x}
  P_{{\rm qg},\, x \ra 0\,}^{\,(n)}(x) \; = \;
  \sum_{\ell\,=\,1}^{n} 
     \, E_{n,\ell}^{\:\rm qg} \: 1/x \,\ln^{\,n-\ell\!} x
  \;+\;
  \sum_{\ell\,=\,0}^{2n}
     \, F_{n,\ell}^{\:\rm qg} \: \ln^{\,2n-\ell\!} x
  \;+\; {\cal O}\,(x\, \ln^{\,a} x)
\; .
\eeq
The coefficients $E_{n,1}^{\:\rm qg}$ of the leading $1/x$ logarithms are known~\cite{Catani:1994sq}, 
as well as those of the highest three sub-dominant $x^{\,0}$ double logarithms, $F_{n,\ell}$, 
for $\ell = 0,\,1,\,2$ at $n=3,\,4$~\cite{Davies:2022ofz}.

Our procedure for the construction of the approximations is analogous
to that for the pure-singlet case in ref.~\cite{Falcioni:2023luc}.
The situation is less favourable here due to the presence of three 
logarithmically enhanced unknown $p = 0$ terms in eq.~(\ref{eq:large-x}). 
This is reflected in a larger uncertainty of the crucial 
$x^{\,-1} \ln x$ coefficient $E_{3,2}^{\:\rm qg}$~in 
eq.~(\ref{eq:small-x}). 
Instead of a `direct fit' of this term, as shown in fig.~1 of 
ref.~\cite{Falcioni:2023luc} for $P_{\rm ps}^{\,(3)}(x)$, we have 
first determined a conservative range for this parameter, 
then constructed 80 approximations for the two boundaries of this range,
and finally selected representatives $P_{{\rm qg,}\,A}^{\,(3)}(x)$ 
and $P_{{\rm qg,}\,B}^{\,(3)}(x)$ that provide the error bands for 
$\nf = 3,\,4,\,5$ light flavours.

\begin{figure}[p]
\vspace{-4mm}
\centerline{\epsfig{file=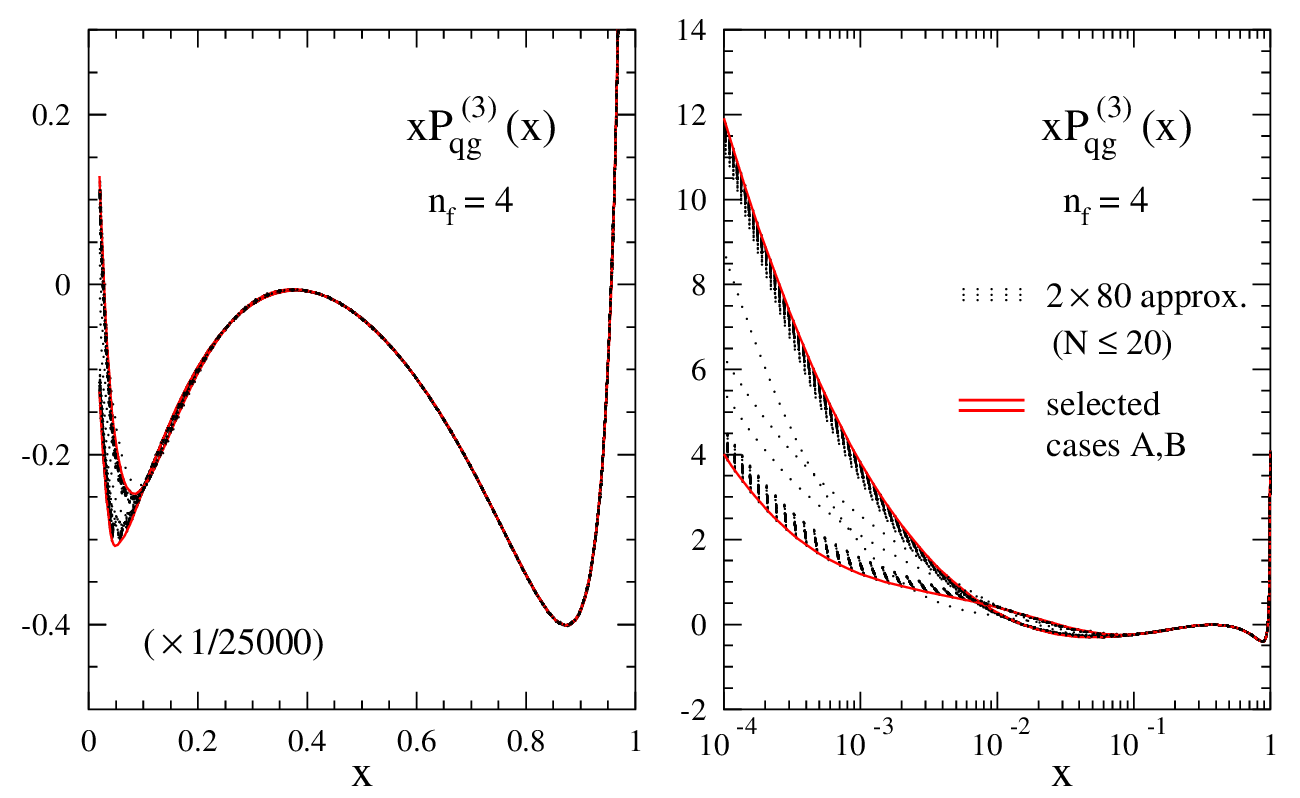,width=16.0cm,angle=0}}
\vspace{-3mm}
\caption{\label{fig:pqg3ab} \small
Two sets of 80 trial functions, one for a large and one for a small 
value of the unknown coefficient of $x^{\,-1} \ln x$, for the 
four-loop (N$^3$LO) contribution to the gluon-to-quark splitting 
function at $\nf = 4$.
The two cases selected for eq.~(\ref{eq:Pqg3A3-nf4}) are shown by 
the solid (red) lines.}
\end{figure}

Using the abbreviations $x_1^{}=1\!-\!x$, $L_1=\ln (1\!-\!x)$ and 
$L_0=\ln x$, the chosen approximations, shown in red in 
fig.~\ref{fig:pqg3ab} for $\nf=4$, are 
\bea
\label{eq:Pqg3A3-nf3}
{\lefteqn{ 
 P_{\rm qg,\,A}^{\,(3)}(\nf=3,x) \; = \; 
 p_{{\rm qg},0}^{\,(\nf=3)}(x) + 
             187500\,\*L_0/x       
           + 826060\,\*x_1/x       
           - 150474
           + 226254\,x\*(2-x)
}}
\nn \\ && \mbox{}
           + 577733\,\*L_0    
           - 180747\,\*L_0^2
           + 95411\,\*L_0^3   
           +  119.8\,\*L_1^3    
           + 7156.3\,\*L_1^2
           + 45790\,\*L_1
           - 95682\,\*L_0\*L_1
\, ,
\nn\\[1mm]
{\lefteqn{ 
 P_{\rm qg,\,B}^{\,(3)}(\nf=3,x) \; = \; 
 p_{{\rm qg},0}^{\,(\nf=3)}(x) + 
             135000\,\*L_0/x
           + 484742\,\*x_1/x
           - 11627
           - 187478\,x\*(2-x)
}}
\nn \\ && \mbox{}
           + 413512\,\*L_0
           - 82500\,\*L_0^2
           + 29987\,\*L_0^3
           -  850.1\,\*L_1^3
           - 11425\,\*L_1^2
           - 75323\,\*L_1
           + 282836\,\*L_0\*L_1
\, ,
\nn \\
\\
\label{eq:Pqg3A3-nf4}
{\lefteqn{ 
 P_{\rm qg,\,A}^{\,(3)}(\nf=4,x) \; = \; 
 p_{{\rm qg},0}^{\,(\nf=4)}(x) + 
             250000\,\*L_0/x       
           + 1089180\,\*x_1/x       
           - 241088
           + 342902\,x\*(2-x)
}}
\nn \\ && \mbox{}
           + 720081\,\*L_0    
           - 247071\,\*L_0^2
           + 126405\,\*L_0^3   
           +  272.4\,\*L_1^3    
           + 10911\,\*L_1^2 
           + 60563\,\*L_1
           - 161448\,\*L_0\*L_1
\, ,
\nn\\[1mm]
{\lefteqn{ 
 P_{\rm qg,\,B}^{\,(3)}(\nf=4,x) \; = \; 
 p_{{\rm qg},0}^{\,(\nf=4)}(x) + 
             180000\,\*L_0/x
           + 634090\,\*x_1/x
           - 55958
           - 208744\,x\*(2-x)
}}
\nn \\ && \mbox{}
           + 501120\,\*L_0
           - 116073\,\*L_0^2
           + 39173\,\*L_0^3
           - 1020.8\,\*L_1^3
           - 13864\,\*L_1^2
           - 100922\,\*L_1
           + 343243\,\*L_0\*L_1
\, ,
\nn\\
\\
\label{eq:Pqg3A3-nf5}
{\lefteqn{ 
 P_{\rm qg,\,A}^{\,(3)}(\nf=5,x) \; = \; 
 p_{{\rm qg},0}^{\,(\nf=5)}(x) + 
             312500\,\*L_0/x       
           + 1345700\,\*x_1/x       
           - 350466
           + 480028\,x\*(2-x)
}}
\nn \\ && \mbox{}
           + 837903\,\*L_0    
           - 315928\,\*L_0^2
           + 157086\,\*L_0^3   
           +  472.7\,\*L_1^3    
           + 15415\,\*L_1^2 
           + 75644\,\*L_1
           - 244869\,\*L_0\*L_1
\, ,
\nn\\[1mm]
{\lefteqn{ 
 P_{\rm qg,\,B}^{\,(3)}(\nf=5,x) \; = \; 
 p_{{\rm qg},0}^{\,(\nf=5)}(x) + 
             225000\,\*L_0/x
           + 776837\,\*x_1/x
           - 119054
           - 209530\,x\*(2-x)
}}
\nn \\ && \mbox{}
           + 564202\,\*L_0
           - 152181\,\*L_0^2
           + 48046\,\*L_0^3
           - 1143.8\,\*L_1^3
           - 15553\,\*L_1^2
           - 126212\,\*L_1
           + 385995\,\*L_0\*L_1
\nn\\
\\[-12mm] \nn
\eea
with the known endpoint contributions \cite{Soar:2009yh,%
Vogt:2010cv,Almasy:2010wn,Catani:1994sq,Davies:2022ofz}
\bea
\label{eq:Pqg30-nf}
{\lefteqn{ \hspn\hspn
 p_{{\rm qg},0}^{\,(\nf)}(x) \: = \: \nf\*\Big\{
 3935.7613\,\*L_0^2/x
-(19.588477-2.7654321\*\nf)\*L_0^6 
+(21.573663+17.244444\*\nf)\*L_0^5
 }}
\nn \\ && \mbox{}
+(-2866.7643+301.22403\*\nf+4.1316872\*\nfs)\*L_0^4
+(1.8518519-0.41152263\*\nf)\*L_1^5
\nn \\[0.5mm] && \mbox{}
+(35.687794-3.5116598\*\nf-0.082304527\*\nfs)\*L_1^4
+(2.8806584+0.82304527\*\nf)\*x_1\*L_1^5
\nn \\ && \mbox{} 
+(-40.511391+5.5418381\*\nf+0.16460905\*\nfs)\*x_1\*L_1^4
\Big\}
\eea
where all coefficients have been rounded to eight significant figures.

These error bands also lead to the following predictions for the 
numerical values of $\gamma_{\,\rm qg}^{\,(3)}(N)$ at $N=22$, with the 
brackets indicating a conservative uncertainty of the last digit(s):
\bea
\label{eq:gqg3N22appr}
  - \gamma_{\,\rm qg}^{\,(3)}(N\!=\!22) \,=\, 
   908.45234(6)\, , \:\:
  1686.02258(8)\, , \:\:
  2721.37166(10)
\quad \mbox{for} \quad \nf\,=\,3,4,5
\; .
\eea

\begin{figure}[p]
\vspace{-2mm}
\centerline{\epsfig{file=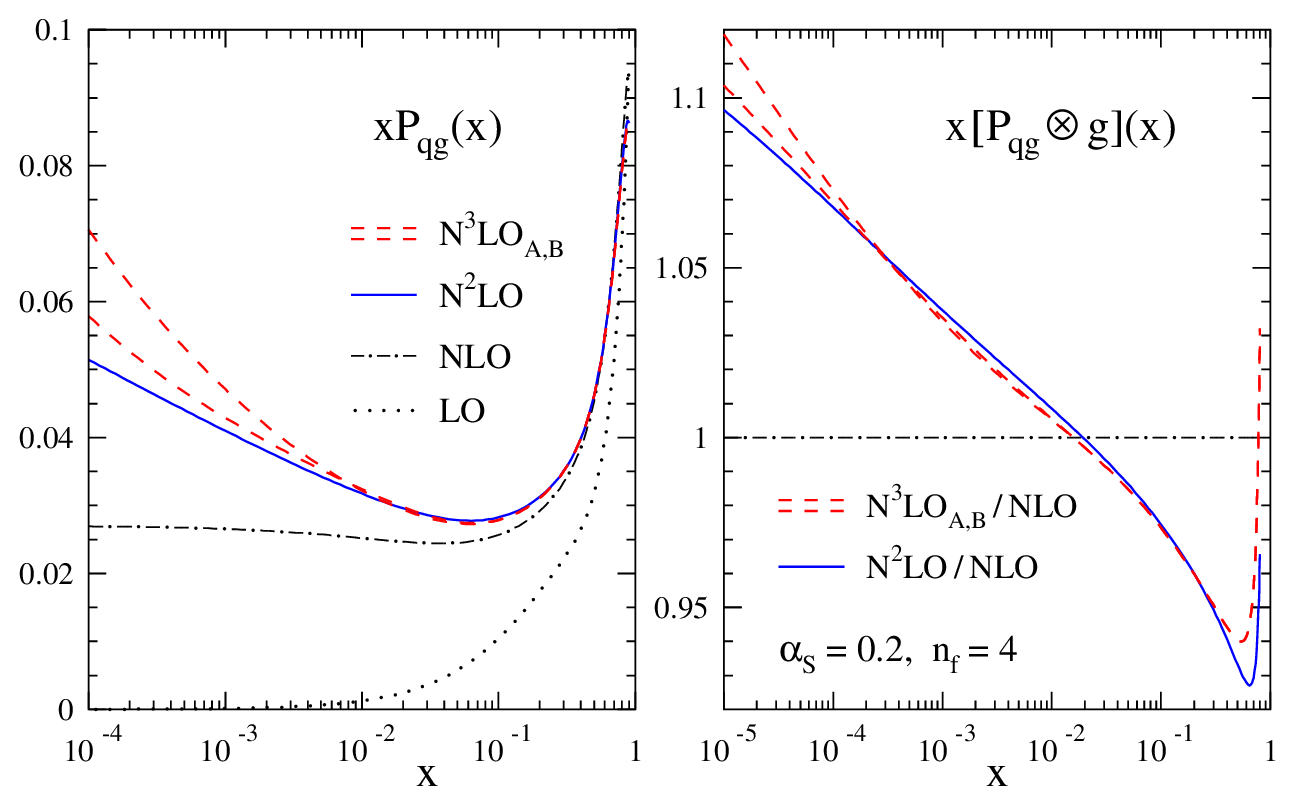,width=16.0cm,angle=0}}
\vspace{-2mm}
\caption{\label{fig:pqgn3lo} \small
Left: The perturbative expansion of the splitting functions 
$P_{\rm qg}$ to N$^3$LO for $\nf = 4$ and $\als = 0.2$, using 
eq.~(\ref{eq:Pqg3A3-nf4}) for the four-loop contribution.
Right: The resulting N$^2$LO and N$^3$LO convolutions (\ref{eq:Mconv})
with the reference gluon distribution (\ref{eq:gluon}), normalized to 
the NLO result.}
\end{figure}

The resulting perturbative expansion of $P_{\rm qg}(x,\als)$ to N$^3$LO
is illustrated in the left panel of fig.~\ref{fig:pqgn3lo} for $\nf = 4$ 
at a standard reference point $\als (\mu_{0}^{\,2}) \, = \, 0.2$ 
corresponding to a scale in the range 
$\mu_{0}^{\,2} \,\simeq\, 25\ldots 50$ GeV$^2$.
The remaining uncertainty due to the approximate nature of
$P_{\rm qg}^{\,(3)}(x)$ is completely unproblematic down to 
$x \simeq 3 \cdot 10^{\,-3}$, but reaches about $\pm 10\%$ at 
$x = 10^{\,-4}$.

This does not mean, however, that the effect of $P_{\rm qg}$ on the scale 
dependence of the singlet quark PDF $q_{\rm s}^{}$ is that uncertain. 
As shown in the right panel of fig.~\ref{fig:pqgn3lo}, the uncertainty of 
the convolution 
\beq
\label{eq:Mconv}
  [ P_{\rm qg} \otimes g ](x) \:=\: 
  \int_x^1 \! \frac{dy}{y} \: P_{\rm qg}(y)\, g\bigg(\frac{x}{y}\bigg)
\eeq
amounts to only about 1\% or less even down to $x = 10^{\,-5}$ for
$\als \, = \, 0.2$ and the sufficiently realistic (order-independent)
reference gluon distribution \cite{Vogt:2004mw}
\beq
  \label{eq:gluon}
  xg(x,\mu_{0}^{\,2}) \; = \; 
  1.6\: x^{\, -0.3} (1-x)^{4.5}\, \left(1 - 0.6\: x^{\, 0.3\,}\right)
\; .
\eeq
Note that in eq.~(\ref{eq:Mconv}) $P_{\rm qg}$ at $y\gsim x$ is 
multiplied by the small $g$ at $x \lsim 1$, while $P_{\rm qg}$ at 
large $y$ is combined with the much larger small-$x$ gluon PDF.
In view of these results we can conclude that the present results 
for $P_{\rm qg}^{\,(3)}$ should be sufficient for a wide range of 
phenomenological applications, even if our error bands were to somewhat 
underestimate its remaining uncertainty at $x \lsim 10^{\,-3}$.


\begin{figure}[btth]
\centerline{\epsfig{file=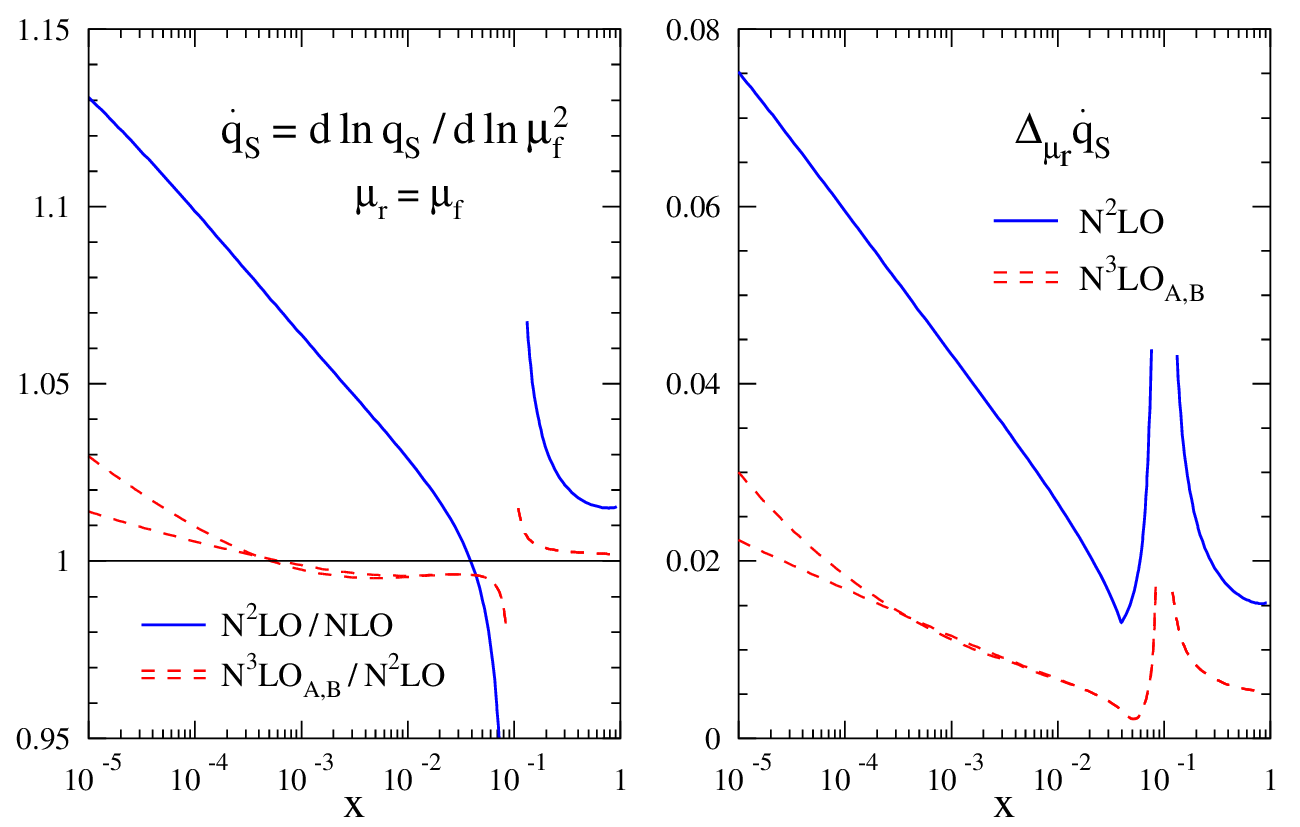,width=16.0cm,angle=0}}
\vspace{-2mm}
\caption{\label{fig:dqsn3lo} \small
Left: The relative N$^2$LO and N$^3$LO corrections to the scale 
derivative of the quark PDF $q_{\rm s}$ at the reference point with 
$\nf=4$ and $\als(\mu_{0}^{\,2}) = 0.2$. 
Right: The renormalization-scale uncertainties of these results, as 
estimated using eq.~(\ref{eq:mur-var}). 
Note that $dq_{\rm s}^{} / d\ln \mu^{\,2}$ changes sign close to
$x=0.1$, which leads to (irrelevant)  singularities in both the 
relative corrections and the relative scale uncertainty 
(\ref{eq:mur-var}).}
\end{figure}

Combining our above results with those of 
refs.~\cite{Moch:2017uml,Falcioni:2023luc}, we are now able to evaluate
the N$^3$LO contributions to the scale derivative (\ref{eq:sgEvol})
of the quark PDF $q_{\rm s}^{}$ at the chosen reference point.
Complementing, as already in ref.~\cite{Vogt:2004mw}, 
eq.~(\ref{eq:gluon}) by
\beq
\label{eq:quark}
  xq_{\rm s}^{}(x,\mu_{0}^{\,2}) \;=\;
  0.6\: x^{\, -0.3} (1-x)^{3.5}\, (1 + 5.0\: x^{\, 0.8\,})
\; ,
\eeq
the relative size of the N$^2$LO and  N$^3$LO contributions to 
$dq_{\rm s}^{} / d\ln \mu^{\,2}$ is shown in the left part of
fig.~\ref{fig:dqsn3lo}.
We see that the N$^3$LO contributions are much smaller than their 
N$^2$LO counterparts. They exceed 1\% only at $x < 10^{\,-4}$, 
and even at $x = 10^{\,-5}$ amount only to about $(2 \pm 1)\%$.

Up to now, we have identified the renormalization scale
$\mu_{\:\!\rm r}^{}$ with the factorization scale $\mu_{\rm f}^{} \equiv \mu$. 
The expansion in eq.~(\ref{eq:Pgamma}) is readily extended to $\mu_{\:\!\rm r}^{} \neq \mu_{\rm f}^{}$, see, e.g., eq.~(2.9) of 
ref.~\cite{vanNeerven:2001pe} for the expression to order $\as(5)$. 
The scale stability of $\dot{q}_{\rm s}^{} \,\equiv\,
d \ln q_{\rm s}^{} / d\ln \mu^{\,2}$ is illustrated in the right part 
of fig.~\ref{fig:dqsn3lo} by the quantity
\beq
\label{eq:mur-var}
 \Delta_{\,\mu_{\:\!\rm r}^{}} \,\dot{q}_{\rm s}^{} \:\: \equiv \:\:
 \frac{
 \max\, [ \,\dot{q}_{\rm s}^{}(x,\mu_{\rm r}^{\,2} 
   = \lambda\, \mu_{\rm f}^{\,2})] 
 \,-\, \min\, [ \,\dot{q}_{\rm s}^{} (x,\mu_{\rm r}^{\,2} 
   = \lambda\, \mu_{\rm f}^{\,2})] }
 { 2 |\, {\rm average}\, [ \,\dot{q}_{\rm s}^{}(x, \mu_r^{\,2} 
   = \lambda\, \mu_{\rm f}^{\,2})]\, | }
\eeq
for the conventional range $\lambda =1/4 \,\dots\, 4$. 
%of renormalization vs.~factorization scale variations.
Also here we see a clear improvement by including the N$^3$LO terms, to
uncertainties below 2\% at $x\gsim 10^{-4}$ and 1\% at $x\gsim 10^{-2}$.

To summarize, we have computed the even moments $N \leq 20$ of the 
four-loop (N$^3$LO) gluon-to-quark splitting function 
$P_{\rm qg}^{\,(3)}(x)$ in the framework of the operator-product 
expansion. 
We have used these results, together with the known constraints 
for $x \ra 1$ and $x \ra 0$, to provide approximations for 
$P_{\rm qg}^{\,(3)}(x)$ that should be sufficient for most
phenomenological applications. 
Further incremental improvements can be obtained by extending the
computations beyond $N = 20$. A larger reduction of the remaining 
uncertainties at small $x$ would be achieved by the determination of 
the hitherto unknown coefficient of the next-to-leading 
$x^{\,-1} \ln x$ small-$x$ contribution to $P_{\rm qg}^{\,(3)}(x)$.

%
% ---------------------------------------------------------------------
%
\vspace*{-1mm}
\subsection*{Acknowledgements}
\vspace*{-1mm}
This work has been supported by 
the Vidi grant 680-47-551 of the Dutch Research Council (NWO),
the UKRI FLF Mr/S03479x/1;
the Consolidated Grant ST/P0000630/1 {\it Particle Physics at the Higgs 
Centre} of the UK Science and Technology Facilities Council (STFC);
the ERC Starting Grant 715049 {\it QCDforfuture};
the Deutsche Forschungsgemeinschaft through the Research Unit FOR 2926, 
{\it Next Generation pQCD for Hadron Structure: Preparing for the EIC}, 
project number 40824754, and DFG grant MO~1801/4-2;
and the STFC Consolidated Grant ST/T000988/1.

%%{\footnotesize
{\small
\addtolength{\baselineskip}{-1.5mm}

%\bibliographystyle{JHEP}
%\bibliography{qgbib}

\providecommand{\href}[2]{#2}\begingroup\raggedright\endgroup

}

\appendix
%
% ---------------------------------------------------------------------
%
\renewcommand{\theequation}{\ref{sec:appA}.\arabic{equation}}
\setcounter{equation}{0}
\section{Mellin moments of $P^{\,(3)}_{\,\rm qg}$}
\label{sec:appA}

Here we present the exact results for the four-loop quark-to-gluon 
anomalous dimensions $\gamma^{\,(3)}_{\,\rm qg}(N)$ at even $N \leq 20$ 
for a general compact simple gauge group.
The numerical values in QCD, i.e., SU($\nc=3$) have been given in 
eq.~(\ref{eq:gqg3-num}) above.
The quadratic Casimir invariants in SU$(\nc)$ are $\ca = \nc$ and 
$\cf = (\ncs-1)/(2\nc)$. The relevant quartic group invariants arise as 
products of two symmetrized traces of four generators $T_r^a$ of the 
fundamental ($R$) or adjoint ($A$) representation, 
\beq
\label{eq:d4def}
  d_{r}^{\,abcd} \; =\; \frac{1}{6}\: {\rm Tr} \, ( \, 
   T_{r}^{a\,} T_{r}^{b\,} T_{r}^{c\,} T_{r}^{d\,}
   + \,\mbox{ five $bcd$ permutations}\, ) 
\eeq
which leads to
\beq
\label{d4FASUn+d4FFSUn}
\dFRAna \;=\;
  \frct{1}{48}\: n_c ( \ncs + 6 ) 
\;\,, \quad
\dFRRna \;=\;
  \frac{1}{96\,\ncs}\: ( \ncf - 6\,\ncs + 18 )
\; .
\eeq
Their QCD values are 
$\dFRAnA \equiv \dfRAnA = 15/16$ and 
$\dFRRnA \equiv \dfRRnA = 5/96$.
{\small{
\bea
\label{eq:GqgN2}
 \lefteqn{ \hspn \gamma_{\,\rm qg}^{\,(3)}(N\!=\!2) \:\equal\:
%%START
%%L %%texGqgN2 =
    \colourcolour{\nf\, \* \cft} \, \* \left( 
      { 16489 \over 729 }
    + { 736 \over 81 }\, \* \zeta_3
    + { 256 \over 9 }\, \* \zeta_4
    - { 320 \over 3 }\, \* \zeta_5 \right)
  + \,\colourcolour{\nf\, \* \ca \* \cfs} \, \*  \left(
    - { 1153727 \over 13122 }
    + { 7108 \over 81 }\, \* \zeta_3
\right. } \nn \\[0.5mm] &&  \mbox{} \left.
    - { 1136 \over 9 }\, \* \zeta_4
    + { 2000 \over 9 }\, \* \zeta_5 \right)
  + \,\colourcolour{\nf\, \* \cas\, \* \cf} \, \* \left( 
      { 763868 \over 6561 }
    - { 12808 \over 27 }\, \* \zeta_3
    + { 2068 \over 9 }\, \* \zeta_4
    + { 40 \over 9 }\, \* \zeta_5 \right)
\nn \\[0.5mm] && \mbox{\hspn}
  + \,\colourcolour{\nf\, \* \cat} \, \*  \left(
    - { 88769 \over 729 }
    + { 31112 \over 81 }\, \* \zeta_3 - 132\, \* \zeta_4
    - { 3560 \over 27 }\, \* \zeta_5 \right)
  + \,\colourcolour{\nf \,\, \* \dfRAna} \, \* \left( 
      { 368 \over 9 }
    - { 992 \over 9 }\, \* \zeta_3
    - { 2560 \over 9 }\, \* \zeta_5 \right)
\nn \\[0.5mm] && \mbox{\hspn}
  + \,\colourcolour{\nfs\, \* \cfs} \, \*  \left(
    - { 110714 \over 6561 }
    - { 272 \over 9 }\, \* \zeta_3
    + { 224 \over 9 }\, \* \zeta_4 \right)
  + \,\colourcolour{\nfs\, \* \ca\, \* \cf} \, \* \left( 
      { 249310 \over 6561 }
    + { 5632 \over 81 }\, \* \zeta_3
    - { 440 \over 9 }\, \* \zeta_4 \right)
\nn \\[0.5mm] && \mbox{\hspn}
  + \,\colourcolour{\nfs\, \* \cas} \, \* \left( 
      { 48625 \over 2187 }
    - { 3572 \over 81 }\, \* \zeta_3 
    + 24\, \* \zeta_4
    + { 160 \over 27 }\, \* \zeta_5 \right)
  + \,\colourcolour{\nfs \,\, \* \dfRRna}\, \*  \left(
    - { 928 \over 9 }
    - { 640 \over 9 }\, \* \zeta_3
    + { 2560 \over 9 }\, \* \zeta_5 \right)
\nn \\[0.5mm] && \mbox{\hspn}
  + \,\colourcolour{\nft\, \* \cf} \, \*  \left(
    - { 8744 \over 2187 }
    + { 128 \over 27 }\, \* \zeta_3 \right)
  + \,\colourcolour{\nft\, \* \ca} \, \* \left( 
      { 3385 \over 2187 }
    - { 176 \over 81 }\, \* \zeta_3 \right)
%%;
%%STOP
\; , \\[2mm]
\label{eq:GqgN4}
 \lefteqn{ \hspn \gamma_{\,\rm qg}^{\,(3)}(N\!=\!4) \:\equal\:
%%START
%%L %%texGqgN4 =
    \colourcolour{\nf\, \* \cft} \, \*  \left(
    - { 8103828487201 \over 104976000000 }
    + { 5100751 \over 81000 }\, \* \zeta_3
    + { 154589 \over 4500 }\, \* \zeta_4
    - { 3158 \over 45 }\, \* \zeta_5 \right)
} \nn \\[0.5mm] &&  \mbox{\hspn}
  + \,\colourcolour{\nf\, \* \ca\, \* \cfs} \, \* \left( 
      { 5121012352507 \over 26244000000 }
    - { 48971263 \over 405000 }\, \* \zeta_3
    - { 143489 \over 750 }\, \* \zeta_4
    + { 951 \over 5 }\, \* \zeta_5 \right)
\nn \\[0.5mm] && \mbox{\hspn}
  + \,\colourcolour{\nf\, \* \cas\, \* \cf} \, \*  \left(
    - { 314624947013 \over 1312200000 }
    - { 2024593 \over 9000 }\, \* \zeta_3
    + { 1674889 \over 4500 }\, \* \zeta_4
    + { 1237 \over 45 }\, \* \zeta_5 \right)
\nn \\[0.5mm] && \mbox{\hspn}
  + \,\colourcolour{\nf\, \* \cat} \, \* \left( 
      { 143199094853 \over 1458000000 }
    + { 11938031 \over 45000 }\, \* \zeta_3
    - { 26904 \over 125 }\, \* \zeta_4
    - { 17917 \over 135 }\, \* \zeta_5 \right)
\nn \\[0.5mm] && \mbox{\hspn}
  + \,\colourcolour{\nf\, \* \dfRAna} \, \*  \left(
    - { 12196 \over 135 }
    - { 81008 \over 225 }\, \* \zeta_3
    + { 15976 \over 45 }\, \* \zeta_5 \right)
  + \,\colourcolour{\nfs\, \* \cfs} \, \* \left( 
      { 37295583467 \over 26244000000 }
    - { 1400864 \over 50625 }\, \* \zeta_3
    + { 707 \over 45 }\, \* \zeta_4 \right)
\nn \\[0.5mm] && \mbox{\hspn}
  + \,\colourcolour{\nfs\, \* \ca\, \* \cf} \, \* \left( 
      { 217239001681 \over 13122000000 }
    + { 4497112 \over 50625 }\, \* \zeta_3
    - { 103669 \over 2250 }\, \* \zeta_4 \right)
  + \,\colourcolour{\nfs\, \* \cas} \, \*  \left(
    - { 7131194093 \over 4374000000 }
\right. \nn \\[0.5mm] &&  \mbox{\hspp}\left.
    - { 12599759 \over 202500 }\, \* \zeta_3
    + { 7591 \over 250 }\, \* \zeta_4
    + { 664 \over 135 }\, \* \zeta_5 \right)
  + \,\colourcolour{\nfs\, \* \dfRRna} \, \*  \left(
    - { 112424 \over 675 }
    - { 2336 \over 75 }\, \* \zeta_3
    + { 10624 \over 45 }\, \* \zeta_5 \right)
\nn \\[0.5mm] && \mbox{\hspn}
  + \,\colourcolour{\nft\, \* \cf} \, \*  \left(
    - { 312015851 \over 364500000 }
    + { 6644 \over 3375 }\, \* \zeta_3 \right)
  + \,\colourcolour{\nft\, \* \ca} \, \* \left( 
      { 338346151 \over 437400000 }
    - { 5192 \over 2025 }\, \* \zeta_3 \right)
%%;
%%STOP
\; , \\[2mm]
\label{eq:GqgN6}
 \lefteqn{ \hspn \gamma_{\,\rm qg}^{\,(3)}(N\!=\!6) \equal
%%START
%%L %%texGqgN6 =
    \colourcolour{\nf\, \* \cft} \, \*  \left(
    - { 2912197809548779709 \over 21613062492000000 }
    + { 1026604067 \over 24310125 }\, \* \zeta_3
    + { 2582141 \over 77175 }\, \* \zeta_4
    + { 1328 \over 147 }\, \* \zeta_5 \right)
} \nn \\[0.5mm] &&  \mbox{\hspn}
  + \,\colourcolour{\nf\, \* \ca\, \* \cfs} \, \* \left( 
      { 324177529264517279 \over 960580555200000 }
    - { 1154450237 \over 9724050 }\, \* \zeta_3
    - { 28952417 \over 154350 }\, \* \zeta_4
    + { 9832 \over 441 }\, \* \zeta_5 \right)
\nn \\[0.5mm] && \mbox{\hspn}
  + \,\colourcolour{\nf\, \* \cas\, \* \cf} \, \*  \left(
    - { 627686002393628869 \over 1729044999360000 }
    - { 6170262713 \over 48620250 }\, \* \zeta_3
    + { 1096679 \over 3087 }\, \* \zeta_4
    + { 47774 \over 441 }\, \* \zeta_5 \right)
\nn \\[0.5mm] && \mbox{\hspn}
  + \,\colourcolour{\nf\, \* \cat} \, \* \left( 
      { 49981299563948069 \over 345808999872000 }
    + { 2383601783 \over 12965400 }\, \* \zeta_3
    - { 689907 \over 3430 }\, \* \zeta_4
    - { 159724 \over 1323 }\, \* \zeta_5 \right)
\nn \\[0.5mm] && \mbox{\hspn}
  + \, \colourcolour{\nf\, \* \dfRAna} \, \*  \left(
    - { 23820479 \over 264600 }
    - { 11627738 \over 33075 }\, \* \zeta_3
    + { 28624 \over 63 }\, \* \zeta_5 \right)
  + \,\colourcolour{\nfs\, \* \cfs} \, \* \left(
      { 1942638296203817 \over 540326562300000 }
\right. \nn \\[0.5mm] &&  \mbox{\hspp}\left.
    - { 113578219 \over 4862025 }\, \* \zeta_3
    + { 28724 \over 2205 }\, \* \zeta_4 \right) 
  + \,\colourcolour{\nfs\, \* \ca\, \* \cf} \, \* \left( 
      { 3261418656515051 \over 216130624920000 }
    + { 122909317 \over 1620675 }\, \* \zeta_3
    - { 600626 \over 15435 }\, \* \zeta_4 \right)
\nn \\[0.5mm] && \mbox{\hspn}
  + \,\colourcolour{\nfs\, \* \cas} \, \*  \left(
    - { 55264268415947 \over 6175160712000 }
    - { 38177677 \over 720300 }\, \* \zeta_3
    + { 133186 \over 5145 }\, \* \zeta_4
    + { 5360 \over 1323 }\, \* \zeta_5 \right)
\nn \\[0.5mm] && \mbox{\hspn}
  + \,\colourcolour{\nfs\, \* \dfRRna} \, \*  \left(
    - { 665983 \over 4725 }
    - { 192736 \over 6615 }\, \* \zeta_3
    + { 85760 \over 441 }\, \* \zeta_5 \right)
  + \,\colourcolour{\nft\, \* \cf} \, \*  \left(
    - { 1262351231147 \over 2572983630000 }
    + { 15268 \over 9261 }\, \* \zeta_3 \right)
\nn \\[0.5mm] && \mbox{\hspn}
  + \,\colourcolour{\nft\, \* \ca} \, \* \left(
      { 34431246007 \over 55135363500 }
    - { 8866 \over 3969 }\, \* \zeta_3 \right) 
%%;
%%STOP
\; , \\[2mm]
\label{eq:GqgN8}
 \lefteqn{ \hspn \gamma_{\,\rm qg}^{\,(3)}(N\!=\!8) \equal
%%START
%%L %%texGqgN8 =
%
         \, \colourcolour{\nf \* \cft} \*  \left( 
        -{990917988466579134913309 \over 5762165963581440000000}
      	+{3183230120837 \over 180033840000} \* \zeta_3 
	+{1481184343 \over 47628000} \* \zeta_4
      	+{398159 \over 5670} \* \zeta_5 \right)
} \nn \\[0.5mm] &&  \mbox{\hspn}
       + \, \colourcolour{\nf \* \cfs \* \ca} \*  \left( 
      	 {173705322188197694847769 \over 411583283112960000000}
	-{5433407245849 \over 60011280000} \* \zeta_3 
	-{417892403 \over 2381400} \* \zeta_4
	-{171271 \over 1620} \* \zeta_5 \right) 
\nn \\[0.5mm] && \mbox{\hspn}
       + \, \colourcolour{\nf \* \cf \* \cas} \*  \left( 
        -{2068466449111368729523 \over 4899800989440000000}
	-{14832708232003 \over 180033840000} \* \zeta_3 
	+{35445949 \over 108000} \* \zeta_4
	+{23311 \over 140} \* \zeta_5 \right) 
\nn \\[0.5mm] && \mbox{\hspn}
       + \, \colourcolour{\nf \* \cat} \*  \left( 
          {336616045154933559893 \over 2099914709760000000}
	 +{697606492357 \over 5143824000} \* \zeta_3 
	 -{26056547 \over 141750} \* \zeta_4
	 -{419459 \over 3780} \* \zeta_5 \right) 
\nn \\[0.5mm] && \mbox{\hspn}
       + \, \colourcolour{\nf \* \dfRAna} \*  \left( 
         -{273996244909 \over 3086294400}
	 -{137047639 \over 396900} \* \zeta_3 
	 +{30298 \over 63} \* \zeta_5 \right) 
       + \, \colourcolour{\nfs \* \cfs} \*  \left( 
          {40554044566337273617 \over 8231665662259200000}
\right. \nn \\[0.5mm] &&  \mbox{\hspp}\left.
	 -{26373124409 \over 1285956000} \* \zeta_3 
	 +{127207 \over 11340} \* \zeta_4 \right) 
       + \, \colourcolour{\nfs \* \cf \* \ca} \*  \left( 
          {36065612407080472327 \over 2939880593664000000}
	 +{5699612263 \over 85730400} \* \zeta_3 
\right. \nn \\[0.5mm] &&  \mbox{\hspp}\left.
	 -{22867163 \over 680400} \* \zeta_4 \right) 
       + \, \colourcolour{\nfs \* \cas} \*  \left( 
         -{4518848403845479427 \over 419982941952000000}
	 -{13175860451 \over 285768000} \* \zeta_3 
	 +{15234743 \over 680400} \* \zeta_4
	 +{818 \over 243} \* \zeta_5 \right) 
\nn \\[0.5mm] && \mbox{\hspn}
       + \, \colourcolour{\nfs \* \dfRRna} \*  \left( 
         -{2023939021 \over 17222625}
	 -{3285578 \over 127575} \* \zeta_3 
	 +{13088 \over 81} \* \zeta_5 \right) 
\nn \\[0.5mm] && \mbox{\hspn}
       + \, \colourcolour{\nft \* \cf} \*  \left( 
         -{47263236736035329 \over 146994029683200000}
	 +{2244679 \over 1530900} \* \zeta_3 \right) 
       + \, \colourcolour{\nft \* \ca} \*  \left( 
          {886247558029 \over 1708914965625}
	 -{35816 \over 18225} \* \zeta_3 \right) 
%
%%;
%%STOP
\; , \\[2mm]
\label{eq:GqgN10}
 \lefteqn{ \hspn \gamma_{\,\rm qg}^{\,(3)}(N\!=\!10) \equal
%%START
%%L %%texGqgN10 =
%
        \, \colourcolour{\nf \* \cft} \*  \left( 
          -{774607400252577911077514539 \over 3916305575568195232500000}
	  -{805380500854 \over 140086918125} \* \zeta_3 
	  +{2705671898 \over 94334625} \* \zeta_4
\right.} \nn \\[0.5mm] &&  \mbox{\hspp}\left.
	  +{1934336 \over 16335} \* \zeta_5 \right)
       + \, \colourcolour{\nf \* \cfs \* \ca} \*  \left( 
           {2957158836400064364217056863 \over 6188729798428752960000000}
	  -{217656420816083 \over 3922433707500} \* \zeta_3 
\right. \nn \\[0.5mm] &&  \mbox{\hspp}\left.
	  -{15390821408 \over 94334625} \* \zeta_4
	  -{1121272 \over 5445} \* \zeta_5 \right) 
       + \, \colourcolour{\nf \* \cf \* \cas} \*  \left( 
          -{6335098018460327267287847261 \over 13924642046464694160000000}
\right. \nn \\[0.5mm] &&  \mbox{\hspp}\left.
	  -{8647744620157 \over 140086918125} \* \zeta_3 
	  +{3177267559 \over 10481625} \* \zeta_4
	  +{312172 \over 1485} \* \zeta_5 \right) 
\nn \\[0.5mm] && \mbox{\hspn}
       + \, \colourcolour{\nf \* \cat} \*  \left( 
           {683009455461651804853128719 \over 4125819865619168640000000}
	  +{1383109617439853 \over 13312502280000} \* \zeta_3 
	  -{5303419507 \over 31444875} \* \zeta_4
\right. \nn \\[0.5mm] &&  \mbox{\hspp}\left.
	  -{5029972 \over 49005} \* \zeta_5 \right) 
       + \, \colourcolour{\nf \* \dfRAna} \*  \left( 
          -{1269333487356283 \over 14522729760000}
	  -{769373679649 \over 2292097500} \* \zeta_3
	  +{7866112 \over 16335} \* \zeta_5 \right) 
\nn \\[0.5mm] && \mbox{\hspn}
       + \, \colourcolour{\nfs \* \cfs} \*  \left( 
           {4356561239541026269442263 \over 745962966774894330000000}
	  -{287267101372 \over 15565213125} \* \zeta_3 
	  +{269096 \over 27225} \* \zeta_4 \right) 
\nn \\[0.5mm] && \mbox{\hspn}
       + \, \colourcolour{\nfs \* \cf \* \ca} \*  \left( 
           {698565087254281295546651 \over 73675354743199440000000}
	  +{2781155392789 \over 46695639375} \* \zeta_3 
	  -{133132504 \over 4492125} \* \zeta_4 \right) 
\nn \\[0.5mm] && \mbox{\hspn}
       + \, \colourcolour{\nfs \* \cas} \*  \left( 
      	  -{261639145927435210838789 \over 24111934279592544000000}
	  -{26791509912217 \over 653738951250} \* \zeta_3
	  +{88731664 \over 4492125} \* \zeta_4
	  +{46688 \over 16335} \* \zeta_5 \right) 
\nn \\[0.5mm] && \mbox{\hspn}
       + \, \colourcolour{\nfs \* \dfRRna} \*  \left( 
          -{181205970624529 \over 1815341220000}
	  -{2257851248 \over 100051875} \* \zeta_3
	  +{747008 \over 5445} \* \zeta_5 \right) 
\\[0.5mm] && \mbox{\hspn}
       + \, \colourcolour{\nft \* \cf} \*  \left( 
          -{2121999454705273487 \over 9785687613471000000}
	  +{53744464 \over 40429125} \* \zeta_3 \right) 
       + \, \colourcolour{\nft \* \ca} \*  \left( 
           {5513232141828253 \over 12708685212300000}
	  -{430256 \over 245025} \* \zeta_3 \right) 
\nn
%
%%;
%%STOP
\; , \\[2mm]
\label{eq:GqgN12}
 \lefteqn{ \hspn \gamma_{\,\rm qg}^{\,(3)}(N\!=\!12) \equal
%%START
%%L %%texGqgN12 =
%
        \, \colourcolour{\nf \* \cft} \*  \left( 
          -{27638793266616676830649446109488607 \over 127379845102027620253476124800000}
	  -{11151211773046551347 \over 399229296500034000} \* \zeta_3
\right.} \nn \\[0.5mm] &&  \mbox{\hspp}\left.
	  +{19574472368767 \over 738574937100} \* \zeta_4
	  +{390599626 \over 2459457} \* \zeta_5 \right)
\nn \\[0.5mm] &&  \mbox{\hspn}
       + \, \colourcolour{\nf \* \cfs \* \ca} \*  \left( 
           {659614817874426101803347003600309841 \over 1273798451020276202534761248000000}
	  -{7188749553100574749 \over 399229296500034000} \* \zeta_3
\right. \nn \\[0.5mm] &&  \mbox{\hspp}\left.
	  -{224595768029407 \over 1477149874200} \* \zeta_4
          -{237402160 \over 819819} \* \zeta_5 \right) 
\nn \\[0.5mm] && \mbox{\hspn}
       + \, \colourcolour{\nf \* \cf \* \cas} \*  \left( 
          -{80001087611088353320752671559280739 \over 168436158812598506120299008000000}
	  -{21516079448853780563 \over 399229296500034000} \* \zeta_3
\right. \nn \\[0.5mm] &&  \mbox{\hspp}\left.
	  +{47713128097 \over 169553475} \* \zeta_4
	  +{1209482915 \over 4918914} \* \zeta_5 \right) 
\nn \\[0.5mm] && \mbox{\hspn}
       + \, \colourcolour{\nf \* \cat} \*  \left( 
           {5604051730074276816849428210811499 \over 33687231762519701224059801600000}
	  +{23670112712533795577 \over 290348579272752000} \* \zeta_3
\right. \nn \\[0.5mm] &&  \mbox{\hspp}\left.
	  -{1902726848671 \over 12207850200} \* \zeta_4
	  -{128078185 \over 1341522} \* \zeta_5 \right) 
\nn \\[0.5mm] && \mbox{\hspn}
       + \, \colourcolour{\nf \* \dfRAna} \*  \left( 
          -{113527295175322043699 \over 1325344317897600000}
	  -{66259676094769 \over 204528444120} \* \zeta_3
	  +{1159790344 \over 2459457} \* \zeta_5 \right) 
\nn \\[0.5mm] && \mbox{\hspn}
       + \, \colourcolour{\nfs \* \cfs} \*  \left( 
           {8636005982934119110563793073 \over 1331293697584716130417152000}
	  -{1459587596116369 \over 86413267640700} \* \zeta_3
	  +{15558056 \over 1756755} \* \zeta_4 \right) 
\nn \\[0.5mm] && \mbox{\hspn}
       + \, \colourcolour{\nfs \* \cf \* \ca} \*  \left( 
           {1130937915286368016463951600099 \over 160415389345331910590760960000}
	  +{1338278114905597 \over 24689505040200} \* \zeta_3
	  -{2830863407 \over 106576470} \* \zeta_4 \right) 
\nn \\[0.5mm] && \mbox{\hspn}
       + \, \colourcolour{\nfs \* \cas} \*  \left( 
      	  -{12276467270462044744218725473 \over 1186504359063105847564800000}
	  -{2493335008939 \over 67335013746} \* \zeta_3
	  +{5661024029 \over 319729410} \* \zeta_4
	  +{26360 \over 10647} \* \zeta_5 \right) 
\nn \\[0.5mm] && \mbox{\hspn}
       + \, \colourcolour{\nfs \* \dfRRna} \*  \left(
          -{14322664324006372519 \over 165668039737200000}
	  -{33040603052 \over 1660133475} \* \zeta_3
	  +{421760 \over 3549} \* \zeta_5 \right) 
\nn \\[0.5mm] && \mbox{\hspn}
       + \, \colourcolour{\nft \* \cf} \*  \left( 
      	  -{16722425084730244813603 \over 115421852616342949797120}
	  +{1754216962 \over 1438782345} \* \zeta_3 \right) 
\nn \\[0.5mm] && \mbox{\hspn}
       + \, \colourcolour{\nft \* \ca} \*  \left( 
           {894866035734231246739 \over 2452265953751458845000}
	  -{15994814 \over 10061415} \* \zeta_3 \right) 
%
%%;
%%STOP
\; , \\[2mm]
\label{eq:GqgN14}
 \lefteqn{ \hspn \gamma_{\,\rm qg}^{\,(3)}(N\!=\!14) \equal
%%START
%%L %%texGqgN14 =
    \colourcolour{\nf\, \* \cft} \, \* \left( 
      -{1819299941960587863099154536601938001 \over 7838759698586315092521607680000000}
      -{7550014892314697177 \over 153549729423090000} \* \zeta_3
\right.} \nn \\[0.5mm] &&  \mbox{\hspp}\left.
      +{13976188277483 \over 568134567000} \* \zeta_4
      +{183968164 \over 945945} \* \zeta_5
\right)
  + \,\colourcolour{\nf\, \* \ca \* \cfs} \, \*  \left(
      {661449743811755399610800628855948383 \over 1205963030551740783464862720000000}
\right. \nn \\[0.5mm] &&  \mbox{\hspp}\left.
      +{71337338061613627 \over 3489766577797500} \* \zeta_3
      -{161710934825771 \over 1136269134000} \* \zeta_4
      -{114514258 \over 315315} \* \zeta_5
\right)
\nn \\[0.5mm] && \mbox{\hspn}
  + \,\colourcolour{\nf\, \* \cas\, \* \cf} \, \* \left( 
      -{22640076494840853303795154046485601 \over 46383193482759260902494720000000}
      -{33002771469947234411 \over 614198917692360000} \* \zeta_3
\right. \nn \\[0.5mm] &&  \mbox{\hspp}\left.
      +{49073876443 \over 186763500} \* \zeta_4
      +{524068451 \over 1891890} \* \zeta_5
\right)
  + \,\colourcolour{\nf\, \* \cat} \, \*  \left(
      {168326075348902041205140436171567 \over 1019410845774928811043840000000}
\right. \nn \\[0.5mm] &&  \mbox{\hspp}\left.
      +{11389523241876713 \over 175310094960000} \* \zeta_3
      -{12677454308339 \over 87405318000} \* \zeta_4
      -{253312408 \over 2837835} \* \zeta_5
\right)
\nn \\[0.5mm] && \mbox{\hspn}
  + \,\colourcolour{\nf \,\, \* \dfRAna} \, \* \left( 
    -{21335213154567939744419 \over 255128781195288000000}
    -{153330754732349591 \over 492146568663750} \* \zeta_3
    +{432531508 \over 945945} \* \zeta_5
\right)
\nn \\[0.5mm] && \mbox{\hspn}
  + \,\colourcolour{\nfs\, \* \cfs} \, \*  \left(
    {129369763652422573392820192451 \over 18610540594934271349766400000}
    -{560072511456047 \over 35792477721000} \* \zeta_3
    +{38014502 \over 4729725} \* \zeta_4
\right)
\nn \\[0.5mm] && \mbox{\hspn}
  + \,\colourcolour{\nfs\, \* \ca\, \* \cf} \, \* \left( 
    {386337614797770939622435479977 \over 77305322471265434837491200000}
    +{2233279925309627 \over 44740597151250} \* \zeta_3
    -{1330332659 \over 55180125} \* \zeta_4
\right)
\nn \\[0.5mm] && \mbox{\hspn}
  + \,\colourcolour{\nfs\, \* \cas} \, \* \left( 
    -{2867710783541657878705947550651 \over 297328163351020903221120000000}
    -{72753816113873311 \over 2147548663260000} \* \zeta_3
    +{2660490407 \over 165540375} \* \zeta_4
\right. \nn \\[0.5mm] &&  \mbox{\hspp}\left.
    +{4808 \over 2205} \* \zeta_5
\right)
  + \,\colourcolour{\nfs \,\, \* \dfRRna}\, \*  \left(
    -{86703311715954607957 \over 1138967773193250000}
    -{30501482088464 \over 1720792198125} \* \zeta_3
    +{76928 \over 735} \* \zeta_5
\right)
\nn \\[0.5mm] && \mbox{\hspn}
  + \,\colourcolour{\nft\, \* \cf} \, \*  \left(
    -{19361358120503651383514249 \over 209159422270739812872000000}
    +{1680448157 \over 1489863375} \* \zeta_3
\right)
\nn \\[0.5mm] && \mbox{\hspn}
  + \,\colourcolour{\nft\, \* \ca} \, \* \left( 
    {20027115402545662837 \over 64941871576982400000}
    -{9525743 \over 6548850} \* \zeta_3
 \right)
%%;
%%STOP
\; , \\[2mm]
\label{eq:GqgN16}
 \lefteqn{ \hspn \gamma_{\,\rm qg}^{\,(3)}(N\!=\!16) \equal
%%START
%%L %%texGqgN16 =
    \colourcolour{\nf\, \* \cft} \, \* \left( 
      -{285309935238500558583542083319026238564164621 \over 1166615122579055289949159356021080064000000}
\right.} \nn \\[0.5mm] &&  \mbox{\hspp}\left.
      -{9810400981630431077238433 \over 140704478549715015936000} \* \zeta_3
      +{175683953534958707 \over 7655986635897600} \* \zeta_4
      +{42585042173 \over 187459272} \* \zeta_5
\right)
\nn \\[0.5mm] &&  \mbox{\hspn}
  + \,\colourcolour{\nf\, \* \ca \* \cfs} \, \*  \left(
      {40459800028988890047781645860994909743510367591 \over 70580214916032845041924141039275343872000000}
      +{673799623136095806925 \over 11370058872704243712} \* \zeta_3
\right. \nn \\[0.5mm] &&  \mbox{\hspp}\left.
      -{25613875001592827 \over 191399665897440} \* \zeta_4
      -{53790127733 \over 124972848} \* \zeta_5
\right)
\nn \\[0.5mm] && \mbox{\hspn}
  + \,\colourcolour{\nf\, \* \cas\, \* \cf} \, \* \left( 
      -{62680864453090912007596475770731383647151899 \over 126036098064344366146293108998705971200000}
\right. \nn \\[0.5mm] &&  \mbox{\hspp}\left.
      -{2744088762432408579029159 \over 46901492849905005312000} \* \zeta_3
      +{12847645023106583 \over 52081541740800} \* \zeta_4
      +{114532560997 \over 374918544} \* \zeta_5
\right)
\nn \\[0.5mm] && \mbox{\hspn}
  + \,\colourcolour{\nf\, \* \cat} \, \*  \left(
      {958218771321200762153969249033062166194424671 \over 5881684576336070420160345086606278656000000}
      +{116754230000061827322827 \over 2233404421424047872000} \* \zeta_3
\right. \nn \\[0.5mm] &&  \mbox{\hspp}\left.
      -{21661099413914861 \over 159499721581200} \* \zeta_4
      -{94309538177 \over 1124755632} \* \zeta_5
\right)
  + \,\colourcolour{\nf \,\, \* \dfRAna} \, \* \left( 
      -{11773700142311179915702121 \over 144584658675391334400000}
\right. \nn \\[0.5mm] &&  \mbox{\hspp}\left.
      -{101169081992668980737 \over 338101880405889600} \* \zeta_3
      +{20687248255 \over 46864818} \* \zeta_5
\right)
\nn \\[0.5mm] && \mbox{\hspn}
  + \,\colourcolour{\nfs\, \* \cfs} \, \*  \left(
      {26913309311367108225950672393546130497983 \over 3693365511043058348609321875419955200000}
      -{34273560040802435057 \over 2342731910584665600} \* \zeta_3
\right. \nn \\[0.5mm] &&  \mbox{\hspp}\left.
      +{920870395 \over 124972848} \* \zeta_4
\right)
  + \,\colourcolour{\nfs\, \* \ca\, \* \cf} \, \* \left( 
      {14127292221783355514812591446014758843649 \over 4308926429550234740044208854656614400000}
\right. \nn \\[0.5mm] &&  \mbox{\hspp}\left.
      +{398082429977183417 \over 8581435569907200} \* \zeta_3
      -{5635299893903 \over 254944609920} \* \zeta_4
\right)
\nn \\[0.5mm] && \mbox{\hspn}
  + \,\colourcolour{\nfs\, \* \cas} \, \* \left( 
      -{752061916161381997694488132348407122609 \over 84488753520592838040082526561894400000}
      -{146648811315659637037 \over 4685463821169331200} \* \zeta_3
\right. \nn \\[0.5mm] &&  \mbox{\hspp}\left.
      +{3756724288103 \over 254944609920} \* \zeta_4
      +{91135 \over 46818} \* \zeta_5
\right)
  + \,\colourcolour{\nfs \,\, \* \dfRRna}\, \*  \left(
      -{405223386945234740808587 \over 5964117170359892544000}
\right. \nn \\[0.5mm] &&  \mbox{\hspp}\left.
      -{14956670451523 \over 938233656360} \* \zeta_3
      +{729080 \over 7803} \* \zeta_5
\right)
  + \,\colourcolour{\nft\, \* \cf} \, \*  \left(
      -{67269559800455311170432078655537 \over 1266065237571321249351886012416000}
\right. \nn \\[0.5mm] &&  \mbox{\hspp}\left.
      +{602706927269 \over 573625372320} \* \zeta_3
\right)
  + \,\colourcolour{\nft\, \* \ca} \, \* \left( 
      {595309329889088262429948677 \over 2276224562921459645214720000}
      -{5825377 \over 4339335} \* \zeta_3
 \right)
%%;
%%STOP
\; , \\[2mm]
\label{eq:GqgN18}
 \lefteqn{ \hspn \gamma_{\,\rm qg}^{\,(3)}(N\!=\!18) \equal
%%START
%%L %%texGqgN18 =
    \colourcolour{\nf\, \* \cft} \, \* \left( 
      -{148012756997831997868202307165553457375623545179496593 \over 579868193437389872966381064860830137193820160000000}
\right.} \nn \\[0.5mm] &&  \mbox{\hspp}\left.
      -{945662253270294339474700289 \over 10533518123321122812382500} \* \zeta_3
      +{5190258175594037257 \over 241325424336496500} \* \zeta_4
      +{641696741936 \over 2487970485} \* \zeta_5
\right)
\nn \\[0.5mm] &&  \mbox{\hspn}
  + \,\colourcolour{\nf\, \* \ca \* \cfs} \, \*  \left(
      {3377593115765589956686838780117931468210706928679589 \over 5684982288601861499670402596674805266606080000000}
\right. \nn \\[0.5mm] &&  \mbox{\hspp}\left.
      +{55713571318360235296073209 \over 567462255801811329960000} \* \zeta_3
      -{30498909550443367027 \over 241325424336496500} \* \zeta_4
      -{1227777154322 \over 2487970485} \* \zeta_5
\right)
\nn \\[0.5mm] && \mbox{\hspn}
  + \,\colourcolour{\nf\, \* \cas\, \* \cf} \, \* \left( 
      -{144518325056948011595975355225600584643601339168691 \over 286637762450514025193465677143267492433920000000}
\right. \nn \\[0.5mm] &&  \mbox{\hspp}\left.
      -{3742668985391472888888028799 \over 56178763324379321666040000} \* \zeta_3
      +{528762966800933 \over 2272206994200} \* \zeta_4
      +{275656349557 \over 829323495} \* \zeta_5
\right)
\nn \\[0.5mm] && \mbox{\hspn}
  + \,\colourcolour{\nf\, \* \cat} \, \*  \left(
      {3783100083290778419858365225493845265835264614203 \over 23605462790042331486520702823563205259264000000}
\right. \nn \\[0.5mm] &&  \mbox{\hspp}\left.
      +{8709621896829003410471989 \over 206003104627212180288000} \* \zeta_3
      -{241960324875770681 \over 1892748426168600} \* \zeta_4
      -{196787959957 \over 2487970485} \* \zeta_5
\right)
\nn \\[0.5mm] && \mbox{\hspn}
  + \,\colourcolour{\nf \,\, \* \dfRAna} \, \* \left( 
      -{1534254855736463458285860487597 \over 19377416686499290875456000000}
      -{263063336077755987407089 \over 915410841198946092000} \* \zeta_3
\right. \nn \\[0.5mm] &&  \mbox{\hspp}\left.
      +{352805410624 \over 829323495} \* \zeta_5
\right)
  + \,\colourcolour{\nfs\, \* \cfs} \, \*  \left(
      {10078764947996863781145704522495961730672265731 \over 1338692846609543524255196843800974552576000000}
\right. \nn \\[0.5mm] &&  \mbox{\hspp}\left.
      -{34109529761049640993 \over 2475998853692454090} \* \zeta_3
      +{84729109246 \over 12439852425} \* \zeta_4
\right)
\nn \\[0.5mm] && \mbox{\hspn}
  + \,\colourcolour{\nfs\, \* \ca\, \* \cf} \, \* \left( 
      {1012878033791555466499189552588051026133120973 \over 551226466250988509987433994506283639296000000}
      +{39384928880780685043 \over 906959287066833000} \* \zeta_3
\right. \nn \\[0.5mm] &&  \mbox{\hspp}\left.
      -{14488003095154 \over 709071588225} \* \zeta_4
\right)
  + \,\colourcolour{\nfs\, \* \cas} \, \* \left( 
      -{558329616172110729967457047573703570713007 \over 68263339473806626623830835232976302080000}
\right. \nn \\[0.5mm] &&  \mbox{\hspp}\left.
      -{144316461298616900803 \over 4951997707384908180} \* \zeta_3
      +{9658443868132 \over 709071588225} \* \zeta_4
      +{154144 \over 87723} \* \zeta_5
\right)
\nn \\[0.5mm] && \mbox{\hspn}
  + \,\colourcolour{\nfs \,\, \* \dfRRna}\, \*  \left(
      -{2376495270782792811304983137 \over 38754833372998581750912000}
      -{4863139453686076 \over 336211891490475} \* \zeta_3
      +{2466304 \over 29241} \* \zeta_5
\right)
\nn \\[0.5mm] && \mbox{\hspn}
  + \,\colourcolour{\nft\, \* \cf} \, \*  \left(
      -{170709087541104871422520948186675189 \over 7547639671882236595469141528787600000}
      +{1256438567596 \over 1276328858805} \* \zeta_3
\right)
\nn \\[0.5mm] && \mbox{\hspn}
  + \,\colourcolour{\nft\, \* \ca} \, \* \left( 
      {169352269869095430064829083933 \over 761807602187670753954439200000}
      -{1643635612 \over 1317160845} \* \zeta_3
 \right)
%%;
%%STOP
\; , \\[2mm]
\label{eq:GqgN20}
 \lefteqn{ \hspn \gamma_{\,\rm qg}^{\,(3)}(N\!=\!20) \equal
%%START
%%L %%texGqgN20 =
    \colourcolour{\nf\, \* \cft} \, \* \left( 
      -{7898888797208836307118697080909594038142131045933137 \over 29841170188590823871954113279387749750325248000000}
\right.} \nn \\[0.5mm] &&  \mbox{\hspp}\left.
      -{24919895802782926655095568033 \over 227671830314589882541320000} \* \zeta_3
      +{8800030293035128897 \over 434668015763982000} \* \zeta_4
      +{321942919546 \over 1120314195} \* \zeta_5
\right)
\nn \\[0.5mm] &&  \mbox{\hspn}
  + \,\colourcolour{\nf\, \* \ca \* \cfs} \, \*  \left(
      {5846396397334222571039848547067118965377809459521073 \over 9547521209923102928561496076923227413192704000000}
\right. \nn \\[0.5mm] &&  \mbox{\hspp}\left.
      +{41608969999126907233361796449 \over 303562440419453176721760000} \* \zeta_3
      -{964457722861951093 \over 8049407699333000} \* \zeta_4
      -{177243557063 \over 320089770} \* \zeta_5
\right)
\nn \\[0.5mm] && \mbox{\hspn}
  + \,\colourcolour{\nf\, \* \cas\, \* \cf} \, \* \left( 
      -{48653145651074538192154470929007688681206494881543387 \over 95475212099231029285614960769232274131927040000000}
\right. \nn \\[0.5mm] &&  \mbox{\hspp}\left.
      -{10028079346191274420177165969 \over 130098188751194218595040000} \* \zeta_3
      +{530885809238315899 \over 2408133051324000} \* \zeta_4
      +{20854255237 \over 58198140} \* \zeta_5
\right)
\nn \\[0.5mm] && \mbox{\hspn}
  + \,\colourcolour{\nf\, \* \cat} \, \*  \left(
      {52740764656642797287118703600606775479812938169233 \over 335000744207828172931982318488534295199744000000}
\right. \nn \\[0.5mm] &&  \mbox{\hspp}\left.
      +{6563217389990563110422497711 \over 191723646580707269508480000} \* \zeta_3
      -{263379457774465111 \over 2178787046436000} \* \zeta_4
      -{111871437401 \over 1493752260} \* \zeta_5
\right)
\nn \\[0.5mm] && \mbox{\hspn}
  + \,\colourcolour{\nf \,\, \* \dfRAna} \, \* \left( 
      -{9934247787263135793881744121471193 \over 129131104798831274394038784000000}
      -{20055818840172257506705213 \over 72622593401783056632000} \* \zeta_3
\right. \nn \\[0.5mm] &&  \mbox{\hspp}\left.
      +{153050440696 \over 373438065} \* \zeta_5
\right)
  + \,\colourcolour{\nfs\, \* \cfs} \, \*  \left(
      {12602646140298382627127034094145561190489093641 \over 1636418690854775628781279322110795868160000000}
\right. \nn \\[0.5mm] &&  \mbox{\hspp}\left.
      -{669887176495794840997 \over 51345159362120373750} \* \zeta_3
      +{61483157 \over 9699690} \* \zeta_4
\right)
\nn \\[0.5mm] && \mbox{\hspn}
  + \,\colourcolour{\nfs\, \* \ca\, \* \cf} \, \* \left( 
      {680809114716993485789591694502123755845386157 \over 1090945793903183752520852881407197245440000000}
      +{234925277032206037451 \over 5744912935621860000} \* \zeta_3
\right. \nn \\[0.5mm] &&  \mbox{\hspp}\left.
      -{3728072666687 \over 196054984125} \* \zeta_4
\right)
  + \,\colourcolour{\nfs\, \* \cas} \, \* \left( 
      -{129237703738338942959554547244734274049443749 \over 17225459903734480302960834969587324928000000}
\right. \nn \\[0.5mm] &&  \mbox{\hspp}\left.
      -{5608977415241450022107 \over 205380637448481495000} \* \zeta_3
      +{4970688711649 \over 392109968250} \* \zeta_4
      +{116488 \over 72765} \* \zeta_5
\right)
\nn \\[0.5mm] && \mbox{\hspn}
  + \,\colourcolour{\nfs \,\, \* \dfRRna}\, \*  \left(
      -{2704920902087866223629472523102433 \over 48424164299561727897764544000000}
      -{108318373596246439 \over 8189020631917125} \* \zeta_3
      +{1863808 \over 24255} \* \zeta_5
\right)
\nn \\[0.5mm] && \mbox{\hspn}
  + \,\colourcolour{\nft\, \* \cf} \, \*  \left(
      {1053102454698939591121970441821583 \over 715504160954056285443058045819200000}
      +{1635389001947 \over 1764494857125} \* \zeta_3
\right)
\nn \\[0.5mm] && \mbox{\hspn}
  + \,\colourcolour{\nft\, \* \ca} \, \* \left( 
      {4557146928634027537304272139 \over 24094839462761828575965150000}
      -{442312337 \over 379053675} \* \zeta_3
 \right)
%%;
%%STOP
\; .
\eea
}}
The expressions (\ref{eq:GqgN2}) -- (\ref{eq:GqgN10}) have also been
computed, in a different manner, in refs.~\cite{Moch:2021qrk,MRUVV-tba}. 
The terms proportional to $\dfRRnA$ and $\dfRAnA$ to $N \leq 16$ have
been obtained before in ref.~\cite{Moch:2018wjh}.
The $\nft$ contributions are known at all $N$ \cite{Davies:2016jie}.

A {\sc Form} file with the results for $\gamma_{\,\rm qg}(N)$ 
at even $N \leq 20$, all partial all-$N$ expressions in the main text, 
and a {\sc Fortran} subroutine of the approximate splitting function 
$P_{\,\rm qg}^{\,(3)}(x)$ have been deposited at the preprint server 
{\tt https://arXiv.org} with the sources of this letter.
They are also available from the authors upon request.

\end{document}